\documentclass{article}
\usepackage{spconf}

\usepackage[utf8]{inputenc}
\usepackage{algorithm,algpseudocodex}
\usepackage{tikz}
\usepackage{amsmath}
\usepackage{amsfonts}
\usepackage{amssymb}
\usepackage{amsthm}
\usepackage{comment}
\usepackage{csquotes}
\usepackage{todonotes}
\usepackage{subfig}
\MakeOuterQuote{"}

\newcommand\R{\mathbb{R}}
\newcommand\N{\mathbb{N}}
\newcommand\E{\mathbb{E}}

\newcommand\Var{\operatorname{Var}}

\usepackage[utf8]{inputenc}
\usepackage[T1]{fontenc}
\usepackage[hidelinks]{hyperref}
\usepackage{url}
\usepackage{booktabs}
\usepackage{amsfonts}
\usepackage{nicefrac}
\usepackage{microtype}
\usepackage{xcolor}
\usepackage{paralist}

\title{$\mathbf{L^2\cdot M = C^2}$\\ Large Language Models Are Covert Channels}

\name{Simen Gaure$^{1}$, Stefanos Koffas$^{2,}$, Stjepan Picek$^{3,2}$, Sondre Rønjom$^{4}$}
 \address{$^1$Norwegian National Security Authority, Norway \\
          $^2$Cybersecurity Group, Delft University of Technology, The Netherlands\\
          $^3$Digital Security Group, Radboud University, The Netherlands \\
          $^4$University of Bergen, Norway
          }

\begin{document}

\maketitle

\begin{abstract}
Large Language Models (LLMs) have gained significant popularity recently. LLMs are susceptible to various attacks but can also improve the security of diverse systems.
However, besides enabling more secure systems, how well do open source LLMs behave as covertext distributions to, e.g., facilitate censorship-resistant communication?
In this paper, we explore open-source LLM-based covert channels. We empirically measure the security vs. capacity of an open-source LLM model (Llama-7B) to assess its performance as a covert channel. Although our results indicate that such channels are not likely to achieve high practical bitrates,
we also show that the chance for an adversary to detect covert communication is low. To ensure our results can be used with the least effort as a general reference, we employ a conceptually simple and concise scheme and only assume public models.
\end{abstract}

\section{Introduction}
\label{sec:introduction}

Model inference depends, among others,
on temperature and internal randomness to create ``liveliness''.
By exchanging internal randomness with binary ciphertexts (assuming a suitable encoding scheme), inference can be considered as a transformation of ciphertexts into specific domain languages (e.g., HTML) of the particular model, resembling format transforming encryption (FTE)~\cite{cryptoeprint:2012/494}. C. Cachin formalizes a model for information-theoretic steganography for covertext distributions~\cite{CACHIN200441}.
While there have been several attempts to embed covert messages into LLM responses~\cite{rrn-stega,imperce,jaggi},
most of them altered the model distribution directly, thus altering the model itself.
As such, we use concepts commonly explored in cryptography.
We note that the idea of connecting machine learning (ML) and cryptography spans decades~\cite{10.1145/1968.1972, 10.5555/647090.713785}, and, more recently, cryptographic concepts were used to provide stronger theoretical foundations for ML security~\cite{9996741, cryptoeprint:2024/162}.

Inspired by Cachin's information-theoretic model for steganography~\cite{CACHIN200441}, we provide a practical and heuristic analysis of partition-based covert channels using covertext distributions defined by LLMs. We extend Cachin's partition-based channel to a setting where the rate is continuously adjusted according to the token distribution of our LLM. Cachin's proposal is conceptually simple and theoretically sound, and its extension to LLM covertext distributions is straightforward. We believe that basing our experiments on the simplicity of the Cachins partition-based approach broadens the impact of our results
Here, ciphertext bits are encoded into token distributions using a partitioning algorithm that adjusts the rate according to varying token distributions. In contrast to recent papers, our objective is to conduct practical experiments to shed light on the relationship between indistinguishability and channel rate for practical LLM covertext distributions.

\section{Preliminaries}
\label{sec:background}

\textbf{Llama 2.}
We use Llama 2
with 7B parameters as it is open-source and ``small'' enough to be run by common users. The text in Llama consists of ``tokens'', words or parts of words. Internally, Llama operates on these tokens, which are represented by integers. In our case, there are \(K=32000\) different tokens. The text generation proceeds by feeding Llama the prompt. The model then produces a set of \(K\) real numbers, \((\ell_i)_{i=1}^K\). These are then scaled by a user-supplied "temperature" \(T\), and converted to a discrete probability distribution,
\begin{equation*}
    q_i = \frac{\exp(\ell_i/T)}{\sum_j^K \exp(\ell_j/T)}.
\end{equation*}

The probabilities are sorted in descending order. Then, a user-supplied cutoff \(C\) is applied, by letting \(c=1+\max \{n \,:\, \sum_i^n q_i < C\}\), and,
\begin{equation*}
    \tilde p_i = \begin{cases}
        q_i &\text{if \(i \le c,\)}\\
        0 & \text{if \(i > c.\)}
    \end{cases}
\end{equation*}

The \(\tilde p_i\)'s are then scaled to have sum~1, \(p_i = \tilde p_i/\sum_j \tilde p_j\) yielding a probability distribution \((\phi_i)\) over the tokens. A token $w_t$ is then sampled from the distribution $\phi_t$. This token, an integer, represents a word or part of a word, which is output from Llama as a text.
The token is also fed back into Llama to change its state, yielding a new probability distribution $\phi_{t+1}$.
Then another token $w_{t+1}$ is sampled from $\phi_{t+1}$.
This continues until an ``end of sequence'' (EOS) token is sampled or a maximum length is reached.\\

\textbf{Chosen Hyperparameters.}
The information in the event that occurs with probability \(p\) is \(-\log_2 p\). The entropy of a probability distribution \((p_i)_{i=1}^k\) is the expected information \(-\sum_{i=1}^k p_i\log_2 p_i\). Entropy can describe the capacity of a communication channel. We will use text generated by Llama as a communication channel. The entropy of the token distributions is then the maximal bit rate we can achieve by encoding bits into token choices.
In Llama, we can control the entropy of the token distributions with the ``temperature'' \(T\) and ``top-p'' \(C\) hyperparameters. Temperatures above \(\approx 1.3\) may result in logorrhea, where the model produces an unintelligible stream of tokens. Our experiments show that we can increase \(T\) above 1 without problems.
We stick to \(T=1.1\), truncating at the \(C=0.95\) quantile.
Our prompt consists of a ``system'' (``Answer like a play by Shakespeare'') and a ``user'' part (``Write a story about a humorous encounter with The Hollow Men''). We are not interested in the actual tokens, only the distribution of probabilities, irrespective of which tokens they refer to. To collect some distributions, using our prompt and a random seed set from the system clock, we let Llama produce 2200 responses., resulting in \(\approx 1000000\) distributions of average entropy \(\approx 1.15\) bits. The token distributions were saved and used for our analysis.

We have removed the distributions with more than 1024 non-zero probabilities (\(\approx 0.6\%\)). Such distributions have on average \(\approx 5500\) non-zero probabilities and \(\approx 7.5\) average entropy.
We assume that a token distribution with many non-zero probabilities does not reflect any reasonable text generation but is possibly an artifact of incomplete model training and that the probabilities have not quite reached numerically zero.
We use the collected distributions for experiments in Section~\ref{sec:experiments}.

In our runs, the distribution's entropy is zero (only a single token of positive probability) for \(\approx 40\%\) of the tokens.
Embedding secrets in these cases would make our scheme more detectable~\cite{zamir2024excuse}.
As a result, we can only use the distributions with higher entropy to encode bits. The mean of the non-zero entropies is \(\approx 1.93\) with median \(\approx 1.55\).
Of course, with a higher entropy distribution, we can encode more than one bit in a single token sample. We look into this in Section~\ref{sec:binarysplitting}.

\section{Encoding Bits by Sampling Tokens}
\label{sec:binarysplitting}

\subsection{Setup}
Rather than using a (pseudo) random generator to choose tokens from the distributions, we will use the message bits. The message bits can be decoded from the generated text by a receiver running the LLM with the same hyperparameters and prompt. We use the generated text as a communication channel in a way that is hard to distinguish from using a random generator so that the communication is covert.

We assume the message bits are (pseudo) random, e.g., a ciphertext. A simple way to ensure this is to XOR each bit with a bit from a random number generator, e.g., AES in counter mode~\cite{Daemen2002}. We assume an adversary knows our model and its parameters but not whether we encode some unknown bits in the tokens or if we sample tokens randomly.
Then, the question is, could an adversary reject the null hypothesis that the tokens have been randomly drawn?

We partition the probabilities into two sets with almost equal sums. To encode a message bit, we use its value to choose between the two sets, and then we draw randomly from the chosen set. The receiver of the text, running the same model, can then recover the message bit by determining which of the two sets the token is drawn from.

Next, we formally describe our scheme. The tokens \(T=\{1,2,\ldots,K\}\) have probabilities \(P = (p_i)_{i=1}^K\). We start out by finding sets \(T_0\) and \(T_1\) with \(T_0 \cap T_1 = \emptyset\), and \(T_0 \cup T_1 = T\), with corresponding probabilites \(P_0\) and \(P_1\) such that \(\sum P_0 \approx \sum P_1\). That is, the tokens have been partitioned into two groups with approximately equal probability sums.  If this cannot be done because no subset of \(P\) sums to something close to \(\frac12\sum P\), we cannot encode a bit in this token choice. If, however, we have found suitable \(T_0\) and \(T_1\), we pick \(T_0\) to encode a \(0\)-bit, and \(T_1\) to encode a \(1\)-bit.
To measure how good the partitioning is, we use the \emph{split entropy}, \(h_s = -(p\log_2 p + (1-p)\log_2 (1-p)\), where \(p = \sum P_0\). As a hyperparameter for the encoding, we choose a \emph{minimum split entropy} \(H_s \le 1\) and disallow partitions with \(h_s < H_s\).

We continue this process with the chosen half of the tokens. That is, if we have chosen \(T_0\), we try to split it into two halves \(T_{00}\) and \(T_{01}\) with approximately equal probability sums \(P_{00}\) and \(P_{01}\). If successful, we encode another bit. We continue this process until we are left with a set \(T_\bot\), which we cannot halve into two almost equal parts. We then draw a token randomly from \(T_\bot\) according to the probabilities in \(P_\bot\).
The pseudocode
is given in Algorithm~\ref{alg:encodebits} and is assumed to replace the random sampling of tokens in the LLM.

\begin{algorithm}
 \footnotesize
 \caption{Use message bits to sample a token}
 \label{alg:encodebits}
 \begin{algorithmic}[0]
  \Function{encodebits}{bits, $Q$, $H_s$} $\rightarrow$ (number of bits encoded, token)
  \LComment{$Q$ is the vector of token probabilities, bits is a vector of message bits}
  \State $P \leftarrow Q$
  \State $i \leftarrow 0$
  \While{\Call{partition}{$P$, $H_s$} $\rightarrow$ $P_0$, $P_1$}
    \State $i \leftarrow i +1$
    \If {bits[i] = 0}
    \State $P \leftarrow P_0$
    \Else
    \State $P \leftarrow P_1$
    \EndIf
  \EndWhile
  \LComment{The partitions are assumed to be indexed by indices of $Q$}
  \State idx $\leftarrow$ $<$Sample index from \(P\)$>$
  \State \Return $i$, idx
  \EndFunction
\end{algorithmic}
\end{algorithm}

As shown in Algorithm~\ref{alg:decodebits}, the decoder uses the reverse process and splits \(T\) into two halves \(T_0\) and \(T_1\), similarly to the encoder. The decoder has received the generated text, knows the chosen token \(w\), and decodes a \(0\)-bit if \(w\in T_0\), and a \(1\)-bit if \(w\in T_1\). It then tries to split again and may decode another bit. If unable to split, it goes on to the next token.

\begin{algorithm}
 \footnotesize
 \caption{Decode message bits from a token}
 \label{alg:decodebits}
 \begin{algorithmic}[0]
  \Function{decodebits}{idx, $Q$, $H_s$} $\rightarrow$ decoded message bits
  \LComment{idx is an index into the token probabilities $Q$}
  \State $P \leftarrow Q$
  \State bits $\leftarrow$ $\emptyset$
  \While{\Call{partition}{$P$, $H_s$} $\rightarrow$ $P_0$, $P_1$}
    \If {idx $\in$ indices($P_0$)}
      \State $<$append 0 to bits$>$
      \State $P \leftarrow P_0$
    \Else
      \State $<$append 1 to bits$>$
      \State $P \leftarrow P_1$
    \EndIf
  \EndWhile
  \State \Return bits
  \EndFunction
 \end{algorithmic}
\end{algorithm}

The split entropy has a relation to the security of the encoding system via \cite[Theorem 2]{CACHIN200441}, stating that if the difference \(|\sum P_1 - \sum P_0|\) is \(\delta\), the security of the scheme is \(\frac{\delta^2}{\log_e 2}\).
With a partition \(P_0\), \(P_1\) with \(\sum P_0 = \frac12(1-\delta)\) and \(\sum P_1 = \frac12(1+\delta)\), the split entropy is a function, \(h(\delta) = -(\frac12(1-\delta)\log_2(\frac12(1-\delta)) +\frac12(1+\delta)\log_2(\frac12(1+\delta)))\). It is easily shown by applying L'H\^opital's rule twice, that \(\lim_{\delta\to 0} (1-h(\delta))/\delta^2 = \frac{1}{2\log_e 2}\). For split entropies \(h_s\) close to 1, we therefore have \(\frac{\delta^2}{\log_e 2} \approx 2(1-h_s)\).  Thus, the security of our encoding in terms of \cite{CACHIN200441} is \(\approx 2(1-H_s)\).

\subsection{Partitioning the \(\mathbf{P}\)}
\label{sec:splitting}

We have used a simple, almost greedy algorithm to split \(P \subset \R^+\) with \(\sum P = 2\delta\).
We choose a \(k \in \N\) so that \(2^k\) is not too large, e.g., \(k=10\).
We sort \(P = (p_i)_i\) in descending order. Then, we find an \(n\) such that \(L = \sum_{i=1}^n p_i \approx \delta-\epsilon\) and \(\sum_{i=1}^{n+k} p_i \approx \delta+\epsilon\), for some \(\epsilon \ge 0\). For every subset \(S\) of \(\{p_i\}_{i=n+1}^{n+k}\), we compute \(s = L + \sum S\), and choose the \(S\) that minimizes \(|s-\delta|\). We let \(P_0 = \{p_i\}_{i=1}^n \cup S\), and \(P_1 = P\setminus P_0\).
With \(p=\frac{1}{2\delta}\sum P_0\), we can compute the split's entropy, \(h_s = -(p\log_2 p + (1-p)\log_2(1-p))\). We deem the split acceptable if \(h_s \ge H_s\).

If we aim for encoding no more than 1~bit in each token, we should use a large \(k\). Since we are partitioning the result again (\(P_0\) or \(P_1\)), this is not necessarily optimal. A near-perfect split may make further splits harder. Our experiments show that lowering \(k\) to \(3\) improves the encoding's performance because then the chosen partition tends to be easier to partition again, leading to more encoded bits in a single token choice.
However, the closer to \(1\) we choose \(H_s\), the larger \(k\) should be. Based on preliminary experiments, we use \(k=\lfloor-\log(1-H_s)-0.5\rfloor\), and clamp \(k\) between \(2\) and \(16\).

Allowing more inaccurate splits will increase the possible encoding bit rate by splitting more probability sets, but it will also increase the possibility of adversaries being able to discover that something is wrong with the generated text. Thus, we need to account for this trade-off.

\section{Detecting Anomalous Token Samples}
\label{sec:tests}

As above, we denote by \(\phi_n\) the discrete distribution from which the \(n\)th token is sampled. We denote its \(K\) probabilities by \((p_k^n)_{k=1}^K\). If we manipulate the token sampling, we must ensure that it still looks like random sampling from the token distributions \(\phi_n\). A manipulation could, e.g., lead to a bias in the probabilities of the tokens that are selected or too many too small and too large probabilities. Such bias could make our covert channel less stealthy and more easily detected.
Here, we describe a simple test that can be used to test if a sequence of tokens has the expected information content.

When we draw a token, it has a specific probability \(p_k^n\), and we denote its information, a random variable, by \(h_n = -\log p_k^n\). The entropy of \(\phi_n\) is denoted by \(H_n = \E(h_n) = -\sum_k p_k^n \log p_k^n\), the expectation of the random variable \(h_n\). We denote by \(D_n = H_n-h_n\) a random variable that measures the deviation from the expected information.
When \(N\) tokens have been sampled, we form the random variable \(D = \frac1N \sum_n D_n\). We have \(\mu = \E(D)=0\) and variance:
\begin{align*}
   \sigma^2 &= \Var D = \Var\left(\frac1N \sum_n H_n-h_n\right)\\
    &= \frac{1}{N^2}\sum_n \Var h_n\\
    &= \frac{1}{N^2}\sum_n \E(h_n^2) - \E(h_n)^2.
\end{align*}

An actual realization of tokens then produces a sequence of tokens with information \(\hat h_n\), and then a \(\hat D = \frac1N\sum_n (H_n-\hat h_n)\), drawn from \(D\).
Now, \(D\) is an average of random variables \(D_n\), and it is no surprise that it is close to being normally distributed.
We can then test how close \(\hat D\) is to \(\E(D)=0\) by computing a \(z\)-score, \(z = |\hat D|/\sigma,\) and compute a \(p\)-value as \(p = 2(1-\Phi(z))\), where \(\Phi\) is the standard normal cdf. The \(p\)-value is the probability that we draw a \(d\) from \(D\) with \(|d| \ge |\hat D|\). A \(p\)-value close to zero is unlikely if tokens are drawn properly from their distributions.

\section{Experimental Results}
\label{sec:experiments}

In our experiments, we encode a stream of random bits as presented in Section~\ref{sec:binarysplitting} and try to detect the presence of the resulting skewed sampling as shown in Section~\ref{sec:tests}.
\begin{figure}[ht]
    \centering
    \subfloat[][Entropy and bit rate]{\includegraphics[width=0.32\textwidth]{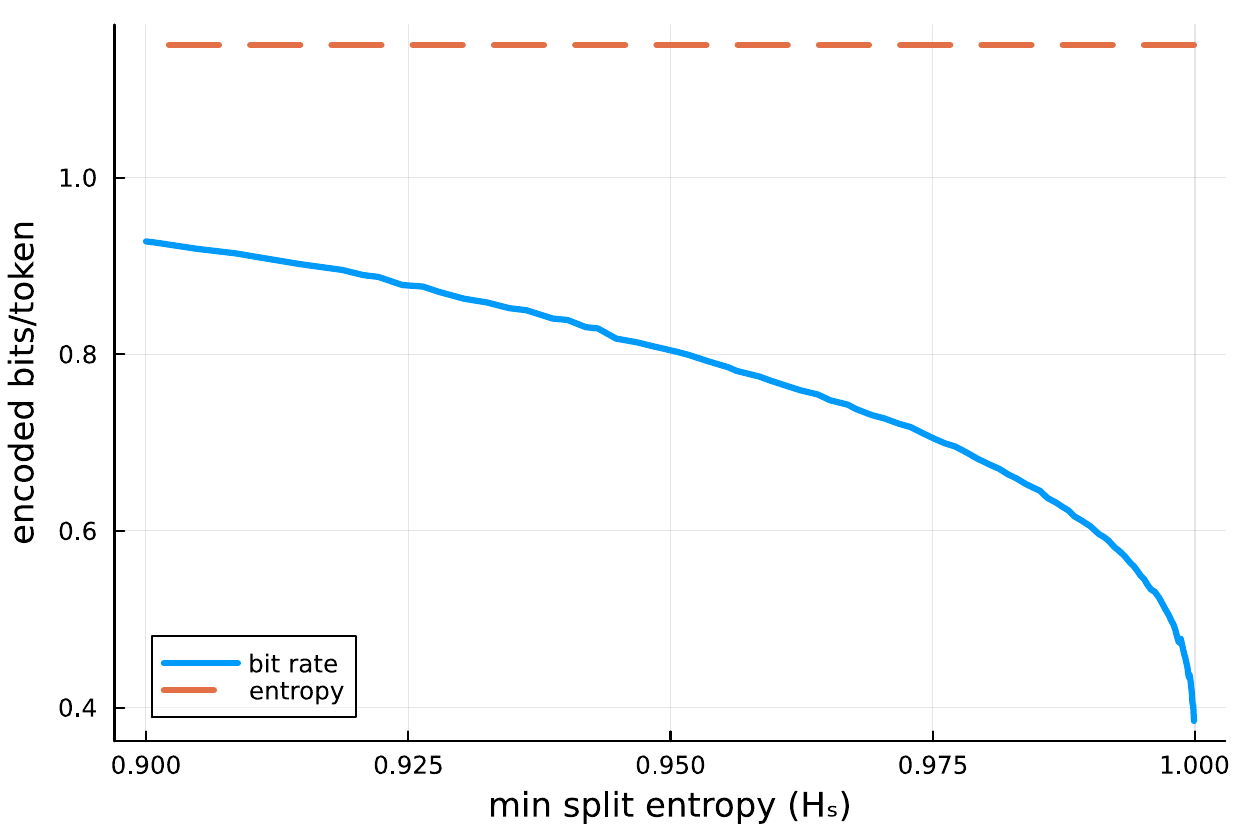}\label{fig:bitrate}}\hfill
    \subfloat[][Detection rate]{\includegraphics[width=0.32\textwidth]{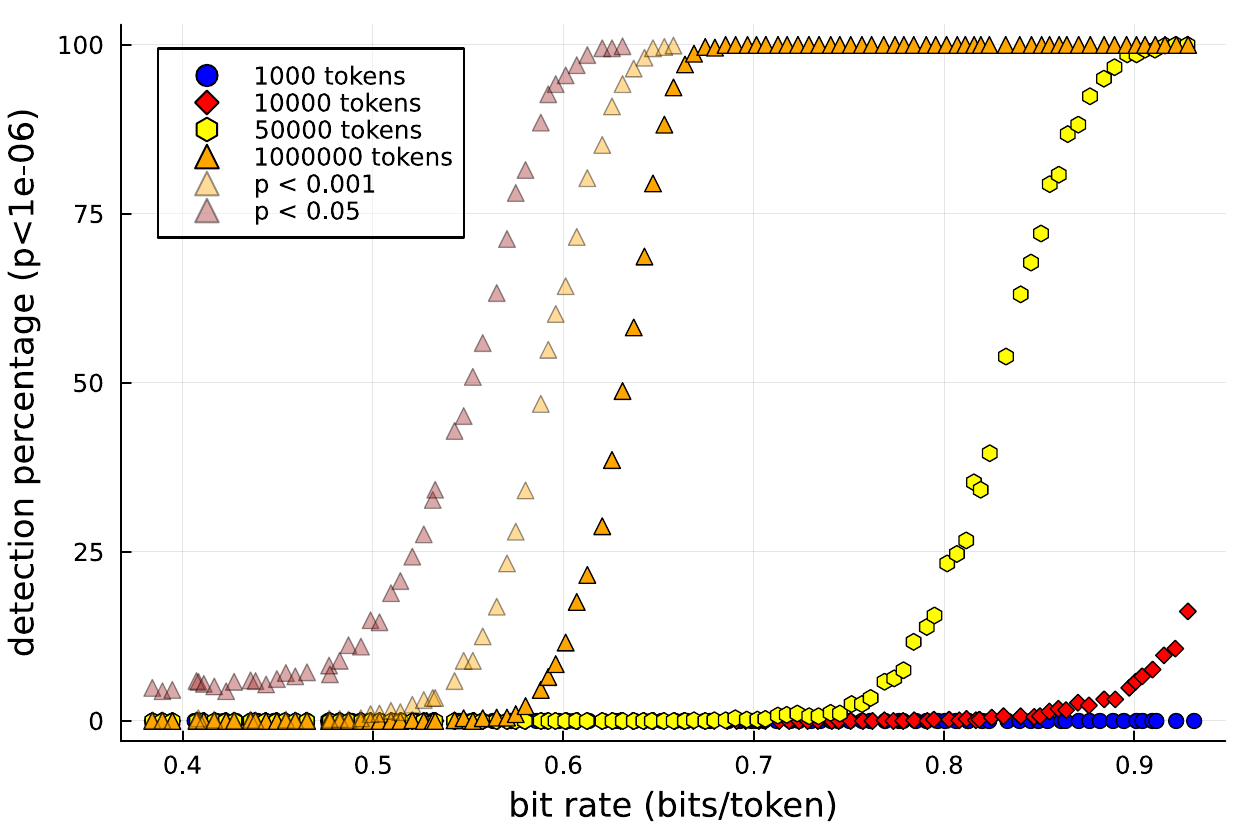}\label{fig:detection-rate}}
    \caption{Bit rate and detection rate}
    \label{fig:bitanddetect}
\end{figure}
In Figure~\ref{fig:bitrate}, the minimum split entropy \(H_s\) varies from \(0.9\) to \(0.9999\). We have plotted the resulting bit rate from encoding random bits in our token distributions. The dashed line is the average entropy of the token distributions. As \(H_s\) approaches \(1\), we see a dramatic drop in bit rate. There are fewer token distributions that can possibly be partitioned with such precision, and therefore, fewer tokens that can be used for encoding bits, resulting in a lower bit rate.

\begin{figure}[ht]
    \centering
    \subfloat[][Theoretically safe bit rates]{\includegraphics[width=0.32\textwidth]{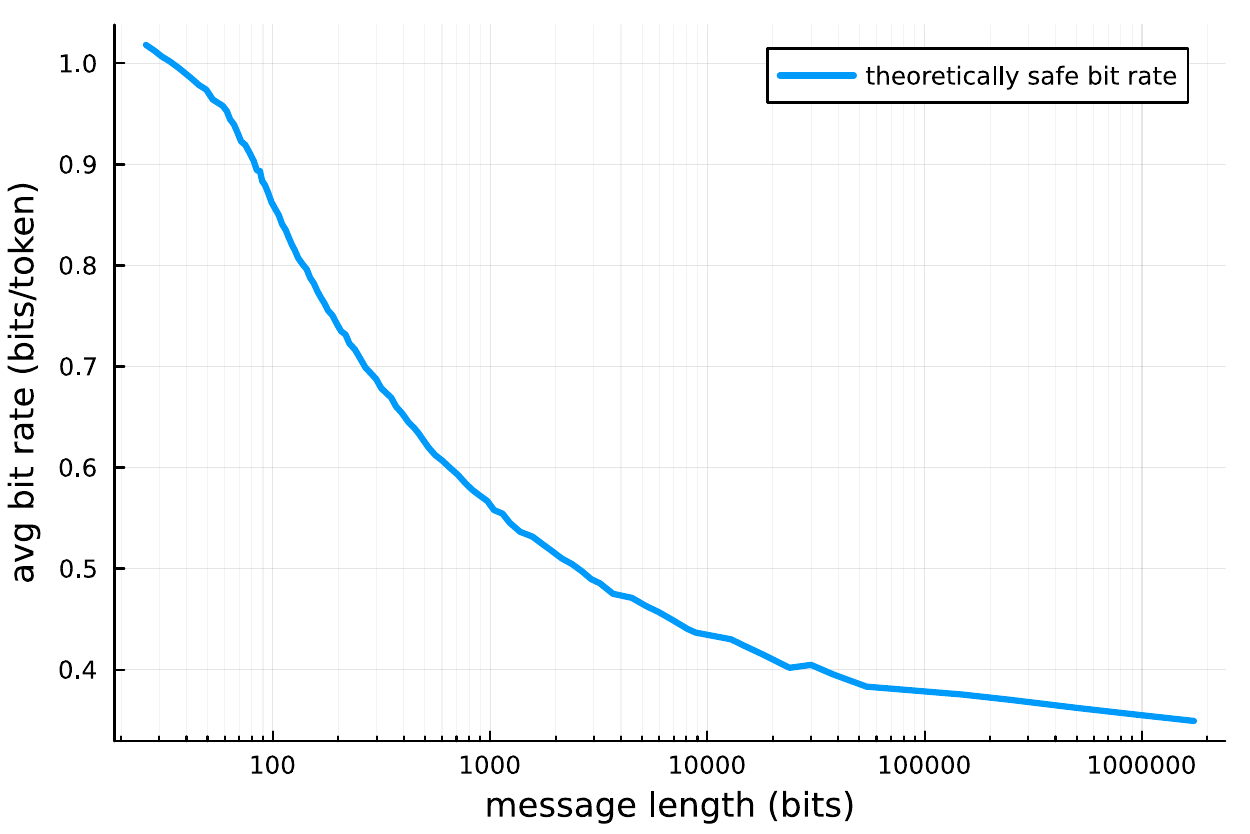}\label{fig:saferate}}\hfill
    \subfloat[][QQ-plots for different bit rates]{\includegraphics[width=0.32\textwidth]{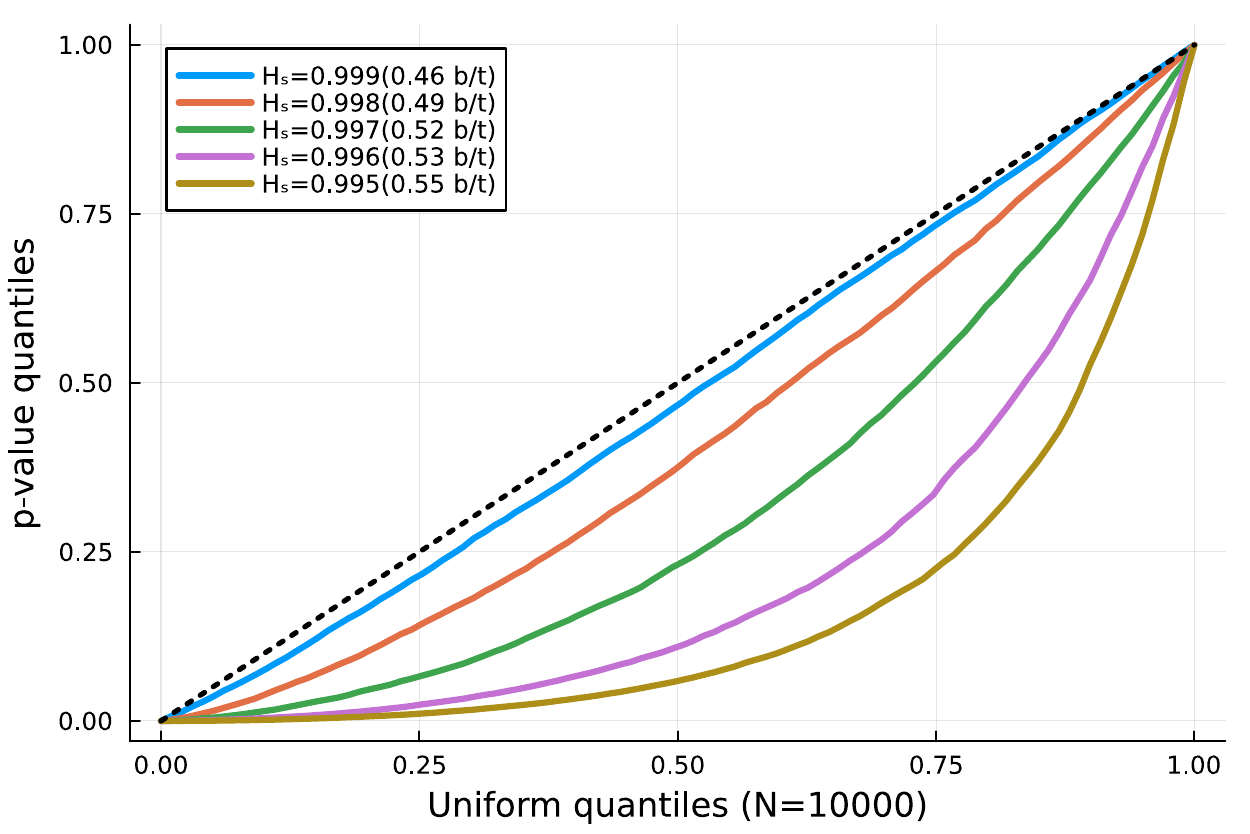}\label{fig:moreqqplots}}
    \caption{Theoretical bounds}
\end{figure}

In Figure~\ref{fig:detection-rate}, we have done 1000 runs of encoding random bitstreams for each \(H_s\), with varying numbers of token distributions drawn from our Llama collection. We have plotted the fraction with the \(p\)-value (from Section~\ref{sec:tests}) below \(10^{-6}\). While this could be considered a somewhat arbitrary significance level, it clearly shows how the bit rate and the number of tokens influence the anomaly detection rate.
For \(10^6\) tokens, we have added significance levels \(0.001\) and \(0.05\) as well.
It is known~\cite{goldreich2017introduction} that distinguishing two finite probability distributions with statistical distance \(\epsilon\) can be done with \(O(\epsilon^{-2})\) samples. So given a limit \(N\) to the number of tokens, we should have \(\epsilon < N^{-1/2}\), for simplicity removing the \(O\).
Our partitioning of the probability distributions changes the probabilities a slight amount, one partition has sum \(p = 0.5-\delta\) whereas the other has sum \(1-p\). This means that the actual probabilities are slightly changed, yielding a statistical distance at most \(\delta\). This translates to a minimum split entropy \(H_s\) and, thus, a theoretically safe bit rate. In Figure~\ref{fig:saferate}, we have plotted these for our set of token distributions.
This is a lower bound for the safe bitrates, though since we have removed a constant factor by replacing \(O(\epsilon^{-2})\) by \(\epsilon^{-2}\), the graph may be horizontally shifted, and should not be taken to be exact.

In Figure~\ref{fig:moreqqplots}, we have used our collection of \(\approx 1000000\) distributions to encode random bits, made \(10000\) p-values for \(H_s \in 0.995..0.999\), and made QQ-plots against a uniform distribution.
We see that the p-values do not become reasonably uniform until the bit rate drops below 0.5 bits/token. But even with a bit rate of 0.46, our \(p\)-values are not entirely uniform.

\section{Conclusion}
\label{sec:conclusion}

It becomes more challenging to recognize if some content is produced by LLMs or humans~\cite{chakraborty2023possibilities, jawahar2020automatic} resulting in research efforts fighting disinformation, such as model watermarking~\cite{christ2023undetectable, munyer2024deeptextmark, wang2024codable, yoo2024advancing, zhang2023watermarks}. However,
the fundamental problem of identifying data produced by a model will likely remain a major challenge in many areas~\cite{jawahar2020automatic, sadasivan2024aigenerated}. Then, if it is computationally infeasible to distinguish synthetic from ``organic'' data, it is computationally infeasible to detect encrypted covert channels based on the same covertext distributions.
We showed that LLMs can be used for covert communication with around 1~bit per word on average bit rate. We conducted extensive experiments on the relationship between covert channel bandwidth and distinguishing probability based on a conceptually simple scheme adopted from the earlier fundamental work on secure covert channels in~\cite{CACHIN200441}. Using a simple and clean scheme allows us to conduct practical experiments that hopefully can be reused by others for both theoretical analysis and comparison with minimal effort. In practice, the detection probability analyzed in this paper is very conservative. It assumes that an adversary knows exactly which LLM is used and its parameters. If this is actually known by a real-world adversary, it is a reasonable assumption that they already know that hidden messages are encoded. If this strong assumption is relaxed, the hidden communication can be virtually impossible to detect.

\bibliographystyle{plain}
\bibliography{biblio}

\appendix

\end{document}